\documentstyle[aps,preprint]{revtex}

\begin{document}
\title{Risk aversion in economic transactions}

\author{C. Anteneodo$^1$,
        C. Tsallis$^1$ and
        A. S. Martinez$^2$}

\address{$^1$
         Centro Brasileiro de Pesquisas F\'{\i}sicas, 
         R. Dr. Xavier Sigaud 150, \\
         22290-180, Rio de Janeiro, RJ, Brazil  \\
         $^2$
         Faculdade de Filosofia, Ci\^encias e Letras de Ribeir\~ao Preto,
         Universidade de S\~ao Paulo, \\Av. Bandeirantes 3900, 
         14040-901 Ribeir\~ao Preto, SP, Brazil. }

\maketitle

\vspace*{1mm} \noindent

\begin{abstract}

Most people are risk-averse  (risk-seeking) when they expect to gain (lose). 
Based on a generalization of ``expected utility theory'' which takes this 
into account, we introduce an automaton mimicking the dynamics of economic operations.
Each operator is characterized by a parameter $q$ which gauges people's  
attitude under risky choices; this index $q$ is in fact the entropic one 
which plays a central role in nonextensive statistical mechanics.
Different  long term patterns of average asset redistribution are observed according 
to the distribution of parameter $q$ (chosen once for ever for each operator) 
and the rules
(e.g., the probabilities involved in the gamble and the indebtedness 
restrictions) governing the values that are exchanged in the transactions.
Analytical and numerical results are discussed in terms of how
the sensitivity to risk affects the dynamics of economic transactions.

\end{abstract}

\vspace{1cm}

People are sensitive to risk,
a characteristic also observed in animals such as rats,
birds and honeybees \cite{animals}.
The usual preference for a sure choice over an alternative of equally
or even more favorable expected value is called {\it risk aversion}.
Actually, most people feel aversion to risk when they expect,
with moderate or high probability, to gain and attraction
to risk when they expect to lose. 
These tendencies are inverted for very low probabilities\cite{at1}.
Certainly, this pattern of attitudes affects most human decisions 
since chance factors are always present, e.g., in medical strategies,
in gambling or in economic transactions.
In particular, in the context of finances,
the attitude of economic operators under risky choices
clearly is one of the main ingredients to be kept in mind
for realistically modeling market dynamics.

In economics, traditionally, the analysis of decision making under
risk was treated through the ``expected utility theory'' (EUT) \cite{eut},
on the assumption that individuals make rational choices.
More precisely, the {\it expected value} $E$, corresponding to the {\it 
prospect}
${\cal P}\equiv(x_1,p_1;\ldots;x_n,p_n)$
such that the outcome $x_i$ (gain if positive; loss if negative) 
occurs with probability $p_i$, is given by
$ E({\cal P}) \,=\,\sum_{i=1}^n \chi(x_i)\,p_i$,
where the weighting function $\chi(x_i)$ monotonically increases with $x_i$.
(Clearly, a statistically fair game corresponds to $\chi(x_i)=x_i$.) 
There are however aspects of risk sensitivity that are not adequately
contemplated within EUT. Such features were exhibited,
through experiments with hypothetical choice problems, by
Kahneman and Tversky\cite{at1}. They then proposed a generalization to EUT
equation within ``prospect theory'' (PT) \cite{at1}:  
$ E({\cal P}) \,=\,\sum_{i=1}^n \chi(x_i)\,\Pi(p_i)$, 
where the weighting function $\Pi(p_i)$ monotonically increases with $p_i$.
Its typical shape (corresponding to the most frequent human attitude) 
is presented in Fig. 1, as sketched by Tversky and collaborators 
\cite{at1,weights} on the ground of experiments and observations.
The weight $\Pi(p)$  is,  basically, concave for low  and convex for high 
probabilities,
with $\Pi(0)=0$ and $\Pi(1)=1$.
Moreover,  most individuals satisfy
(i) $\Pi(p)+\Pi(1-p)\le1$ (the equality holds for $p=0,1$) ({\em 
subcertainty}),
(ii) $\Pi(p\,s)/\Pi(p)\leq\Pi(p\,r\,s)/\Pi(p\,r)$ for $0<p,r,s\leq 1$ ({\em 
subproportionality}),
(iii) $\Pi(p)<(>)p$ for high (low) probabilities ({\em 
under(over)weighting}), and,
(iv) for very  low probabilities,  $\Pi(p\,r)>r\Pi(p)$ for $0<r<1$ ({\em 
subadditivity}).
The following functional forms have been proposed \cite{ct1} in the context 
of nonextensive statistical mechanics \cite{ct2}:
$ \Pi(p)= p^q,\;(q \in \Re)$ and 
$ \Pi(p)= p^q/( p^q + (1-p)^q ) $, 
usually referred to as {\it escort probability}. 
Other functional forms are also available in the literature \cite{shape}, 
such as
$ \Pi(p)= p^q / [p^q + (1-p)^q]^{1/q}$ and 
$ \Pi(p)= p^q / [p^q+ A(1-p)^q] $,  
where $A>0$. Clearly, $A=1$ recovers the escort probability and  
for appropriate choices of $(q,A)$, the latter expression can satisfy  
all the properties detailed above.
In all these cases, each individual is characterized by a set of 
parameters which yields a particular $\Pi(p)$ representing
the subjective processing that the individual makes of
known probabilities $p$  in a chance game. 

More recently, PT was generalized \cite{cpt} using a rank dependent 
or cumulative representation where the ``decision weight" 
multiplying the value of each outcome is distinguished from the 
probability weight. This interesting generalization is however 
irrelevant for the present study. Indeed, we will deal here with 
simple prospects with a single positive outcome, hence both 
versions coincide.

In the present work we investigate the consequences of risk averse 
attitudes in the dynamics of economic operations. In order to do so we apply 
methods of statistical physics. This strategy has proved to be very useful 
in several previous works \cite{physics} (see also \cite{econophys} for 
general discussions on the application of statistical physics methods in economy). 
Here we introduce an automaton simulating monetary transactions among 
operators with different attitudes under risky choices.  
Elementary operations are of the standard type used in hypothetical choice 
problems that exhibit risk aversion \cite{at1}. 
By following  the time evolution of the asset position of the operators, it will be 
possible to conclude on the consequences of each particular attitude.

We will restrict our study to the regime of moderate and high probabilities 
where most people are risk averse for gains. In  this regime, human behavior 
can be satisfactorily described by the weighting function $\Pi(p)=p^q$. 
This expression, which has a simpler form than other weights describing the full 
domain, is the one adopted in the present work. 
Furthermore, since we will focus on probabilistic factors, we assume that all 
individuals have the simplest utility function, namely $\chi(x)= x$.
Adopting these choices and, additionally, assuming that the probability 
weight $\Pi(p)$ is the same for gains and losses, a unique parameter ($q$) 
characterizes each individual according to the attitude under risky choices.
An ideally  rational individual has $q=1$ while most individuals
``feel'' probabilities with $q>1$.

The  present  automaton simulates  monetary 
transactions among  $N$ operators. 
We assume for simplicity that in each elementary transaction 
two agents participate.  
One of them proposes to the other a choice between two alternative ways 
of either paying money or of receiving money.  
As an illustration of the former case, the proponent typically asks:
``What do you prefer: to receive a {\it certain} quantity $X$ or to play
a game where you receive $Y$ with probability $P$, such that $PY=X$, 
and nothing with probability $1-P$ ?''. 
More precisely, the alternative choices are: 1) 
a {\it certain} prospect ${\cal P}=(\pm \alpha\, S,1)$,
with $\alpha>0$, and 2) a {\it risky} one ${\cal P}=(\pm S,P;0,1-P)$, where 
$S$ is a positive quota and the $+/-$ apply to the cases when the proponent 
receives/pays. 
The value of $\alpha$ depends on the psychological factors $q$ of
both operators in a way that will be defined below for each one of the 
models conceived. 
The also possible case where the choice is between a {\it fixed} 
certain prospect ${\cal P}=(\pm S,1)$ and a {\it variable} risky one 
${\cal P}=(\pm \alpha S,P;0,1-P)$ will not be analyzed here
since it yields quite analogous results.
Along the dynamics, the probability $P$ and the quota $S$, 
the two parameters characterizing the elementary operations, are kept fixed. 
Clearly, the present games are not the kind of operations that actually occur 
in a financial market. However in the sense of the theory of financial decisions, 
they illustrate the risk aversion phenomenon.

Let us consider $N$ agents with values of $q$ uniformly distributed
in the interval $[Q_1,Q_2]$.
For simplicity we take $q_k=Q_1+(Q_2-Q_1)(k-1)/(N-1)$ with $k=1,\ldots,N$.
Each operator $k$ has an initial amount of money $M_k(0)$.
In each time step, a transaction between the agents of a randomly chosen
ordered pair ($i$,$j$) occurs.
The exchanged quantity is taken positive if the proponent ($i$)
is the one who receives the money and negative otherwise.
Whether the exchanged quantity is positive ($i$ receives the money) or 
negative ($j$ receives the money), it is randomly drawn  at each  step 
of the dynamics.

Restrictions on the level of indebtedness of the operators can be imposed. 
We consider three cases (NR, OR and PR): 
In case NR, there are {\it no restraints} and operators can be indebted 
without limit; 
in OR, there are {\it opportunistic} restraints and agents can also operate 
indefinitely except that they do not pay when they would have to do so 
if at a given step of the dynamics their asset become less than a minimal 
quantity $M^*$ (i.e., operators can become swindlers occasionally);
finally, in PR, agents become {\it permanently} forbidden to trade from 
the instant when their assets fall below the threshold $M^*$. 
This threshold introduces nonlinearity into the problem. 

For each one of these three cases, other variations (A, S and C)
were considered following different definitions of $\alpha$:
in A (alter-referential), operator $i$, the proponent, somehow knows
the psychology of $j$ ($q_j$);
in S (self-referential), operator $i$ ignores $q_j$ and hence
attributes to $j$ her/his own value of $q$; finally,
in C (consensual), operators $i$ and $j$ act by consensus. 

In variations A and S, the proponent, $i$, will present,
according to the hypothesis made on the $q$ of $j$ ($q_j^{hyp}$ taken to be  
$q_j$ for model A and $q_i$ for model S), 
a certain alternative which $i$ considers the worst
between the one which has the same standard expected value as the game
($\pm P S$) and the one that $i$ believes to have for $j$
the same expectation as the game
($\pm P^{q_j^{hyp}} S$).
This opportunistic behavior can be expressed through the factor
$\alpha$, as presented below: 

Case A:  $\;\;\;\alpha= \min(\max)\{P,P^{q_j}\}$ if $i$ pays(takes).

Case S: $\;\;\;\alpha= \min(\max)\{P,P^{q_i}\}$ if $i$ pays(takes).

Note that, in a statistically  fair game, it should be $\alpha=P$
so that both prospects (the certain and the risky ones)
would have the same standard expected value.
On the other hand, agent $j$ will choose the risky prospect
either 1) if $P^{q_j}>\alpha$ when $i$ pays or
2) if $P^{q_j}<\alpha$ when $i$ takes or 3)
in 50\% of the cases if $P^{q_j}=\alpha$.

Case C: In this instance, there is not a proponent and both
players have a symmetric role arriving to a consensus.
Hence, let us consider without loss of generality,
the case when $i$ pays and $j$ takes.
If $q_i<q_j$ there is agreement for
the risky alternative, otherwise, the agreement is for the certain choice
with $\alpha$ an intermediate value between $P^{q_i}$ and $P^{q_j}$, for
instance, as we will adopt, $\alpha=(P^{q_i}+P^{q_j})/2$. 

Combining all these possibilities, we have a total of nine models: 
(NR,A), (NR,S),$\ldots$,(PR,C). 

Let us discuss first the models of the type (NR,*), 
for which there is no indebtedness restriction. 
The amount of money of operator $i$ at time $t$, $M_i(t)$, 
{\it in average} over a large number of realizations (histories),
is given by

\begin{equation} \label{MM0}
\overline{M_i}(t)=\overline{M_i}(0) + \frac{t}{N}\,S\, V_i 
\;\;\;\;\;\;\forall{i,t},
\end{equation}
where $V_i = \frac{1}{N}\sum_j (G_{ij}-G_{ji})$, 
such that the $G$ matrix has non negative elements.
For instance, for case (NR,C), explicitly one has

\begin{equation}
\label{Gij} G^{C}_{ij}\;=\;
\left\{
\begin{array}{cc}
\frac{1}{2}( P^{q_i}+P^{q_j} )    &\mbox{ $j<i$ },\\
0                                 &\mbox{ $j=i$ },\\
P                                 &\mbox{ $j>i$ }.
\end{array}
\right.
\end{equation}

The continuum approximation of $V_i$ for all models (NR,*),
considering uniform distribution of parameter $q$ in the interval
$[Q_1,Q_2]$, results

\begin{eqnarray} \nonumber
&&V(q)\,\simeq {\frac{1}{Q_2-Q_1} }\times\\ \nonumber         
&&\left\{
\begin{array}{ll}
\frac{1}{4}\left({ (Q_2+Q_1-2)P+ 
                     \frac{2P-P^{Q_1}-P^{Q_2}}{\ln{P}}+
                      (Q_2-Q_1)(P-P^q)\mbox{sign}(1-q)} \right)
                           & \mbox{  (A)}\\
\frac{1}{2}(1-q)P+ \frac{1}{2}\frac{P^q-P}{\ln{P}}-\frac{1}{2}(P-P^q) \times

 \biggl\{     \begin{array}{ll}{(q-Q_1)} &\mbox{ if $q<1$}\\
                         { (q-Q_2)} &\mbox{ otherwise}
        \end{array}
                   &\mbox{  (S)}\\
{ (Q_2+Q_1-2q)(P-\frac{1}{2}P^q)+\frac{2P^q-P^{Q_1}-P^{Q_2}}{2\ln{P}}}
                            &\mbox{  (C)}   
\end{array}
\right.  \\ 
\label{Vq}
\end{eqnarray}

$V(q)$ vs $q$ for models (NR,*) is illustrated in Fig. 2.
$V(q)$ rules the average evolution of assets, being
\begin{equation} \label{MMc}
\overline{M}(q,t)=\overline{M}(q,0)+\frac{t}{N}SV(q)\;\;\;\;\forall q,t.
\end{equation} 
From Fig. 2a corresponding to a more realistic 
distribution of $q$ than Fig. 2b 
(since about 75\% of the people are risk averse when high probabilities are involved),
note that for (NR,A) (where the proponent knows the psychology 
of the other) the maximum emerges at $q=1$ (the rational player) while
for (NR,S) (where the proponent acts following the own psychology) 
we observe the emergence of a minimum at $q=1$ and of two maxima on both 
sides of $q=1$  being the absolute maximum the one for $q>1$ 
(agents who are conservative for gains). 
Case (NR,C) (consensus) corresponds to an intermediate situation
where the maximum asset increase occurs for $q$ above 1.

For models (OR,*) and (PR,*) the average time evolution of $M$ is not linear with $t$.
In these cases, the evolution follows Eq. (\ref{MM0})
(or its continuous version Eq. ({\ref{MMc}}))
up to time $\tau$ when some agent's asset falls below the threshold $M^*$.
In models (OR,*), from that instant on,
the discrete Markovian automaton can be described, in  average, by

\begin{equation} \label{MM}
\overline{M_i}(t+1)= \overline{M_i}(t) +
\frac{S}{N^2} \sum_j (  \overline{H_j}G_{ij}-\overline{H_i} G_{ji} )
\;\;\;\forall i, \; t>\tau,
\end{equation}
where $H_k\equiv H(M_k(t)-M^*)$ is a Heaviside function.
$\overline{M_i}(t)$ evolves to a non  trivial
steady state whose extrema coincid with those of $V_i$.
This steady state $M^{SS}_k$ is approximately of the form
$ M^{SS}_K = a/(b+V^{max}-V_k),\;\forall k $, 
where $a, b>0$. Its continuous version is the long time solution
of the nonlinear diffusion equation
$\partial_t M(q,t)=\partial_q( M(q,t)\partial_q V(q) )+ a\partial_{qq}\ln M(q,t)$. 
In fact, $(-V(q))$ acts as the potential of an  effective drift which rules 
the dynamics.
In Fig. 3, $M(q,t)$, averaged over several realizations,
vs. $q$, is illustrated for models (OR,*). 
The maxima of $\overline{M}$, $\overline{M}_{max}$,
and its corresponding $q$, $q_{max}$, are
plotted as a function of time $t$ in Figs. 3b and 3c to 
control the stability of the resulting state.
The plots are invariant when all monetary amounts
(i.e, $M_o$, $M^*$ and  $S$) are multiplied by a common factor.
As either $M^*/M_o$ or $S/M_o$ increase, the steady state broadens.
Increasing $S$ shortens the time scale and increases the amplitude of 
fluctuations.
Although in the example $M(q,0)=M_o$, the steady state does not
depend on the initial distribution of money.

For models (PR,*), the dynamics follows a different evolution.
$\overline{M}(q,t)$  evolves to a
Dirac $\delta$ function centered at $q=1$ or at the boundary
closer to $q=1$ (see  Fig. 4).
As for the (OR,*) models, the stability is  controlled by watching $\overline{M}_{max}$
and $q_{max}$ vs  t.
Note that in all cases $q_{max}\rightarrow 1$, although for case
(PR,S) the convergence is  slower.
Therefore, with this kind of indebtedness restraint, the more rational 
operator wins at the long  term. Here the final state does not
depend on the distribution of $q$, which  will just generate a different 
$V(q)$, as soon as it contains $q=1$ (if not, the boundary 
closer to $q=1$ wins).
On the other hand, the final state depends on the initial distribution of 
assets once some individuals may have assets below the threshold and are 
forbidden to play from the beginning.

On average, the rational player (with $q=1$) wins from every other
($G_{uj}-G_{ju}\geq 0$, $\forall j$ and $q_u=1$).
That is why with  restraints of type PR such player is the long term winner.
Operators with extreme values of $q$ are long term losers. 
With indebtedness restrictions of kind PR, they have to abandon the game. 
With restrictions of kind OR (cheating is not forbidden), they are allowed to 
remain in the game, 
but their assets keep fluctuating close to the threshold $M^*$. 
A non trivial steady state appears in this case.
Operators with  $q_k$ a bit above one (a bit risk averse for gains)
lose from some neighbors but win more in the total sum,
i.e., $\sum_j (G_{kj}-G_{jk})\geq 0$), as soon as those with  
extreme values of $q$ remain in the game.
Consequently, the maximum of the steady state is located at $q$ above one 
(primacy  of the conservatives). 
In case (OR,*) the initial distribution of parameter $q$ will 
affect the form of the
function $V$ and therefore the shape of the steady state
governed by the effective potential $(-V)$. 
On the other hand, the steady state does not depend on the initial
distribution of money.

In conclusion,
the type of conditions limiting indebtedness are critical for defining
the nature of the long term evolution, i.e., existence or not of
a nontrivial steady state. The details of this steady state depend,  
among other factors, on the distribution of the parameter $q$ of the operators. 
One also observes that the final state is invariant under initial
redistribution of money.  
Paradoxically enough, some level of cheating avoids extreme wealth 
inequality to become the stationary state.  
However, one must keep in mind that the
distribution of $q$ is kept fixed along the dynamics and, therefore,
the psychological effect of asset position is not taken into account
in the present model. Such dynamics would provide an improved, 
more realistic model. 
In fact, a model which, in addition to this 
learning--from--experience dynamics, 
would use a weight such as that of Fig. 1 simultaneously with an 
appropriate nonlinear utility function, might very well constitute 
a quite realistic frame for taking into account the well known human 
risk aversion in the context of collective trading.

We thank R. Maynard for useful discussions at the preliminary
stages of these ideas.
We acknowledge Brazilian agencies CNPq, FAPERJ and PRONEX for
financial support.

\section*{Captions for figures}
\mbox{} \\[-5mm] \noindent
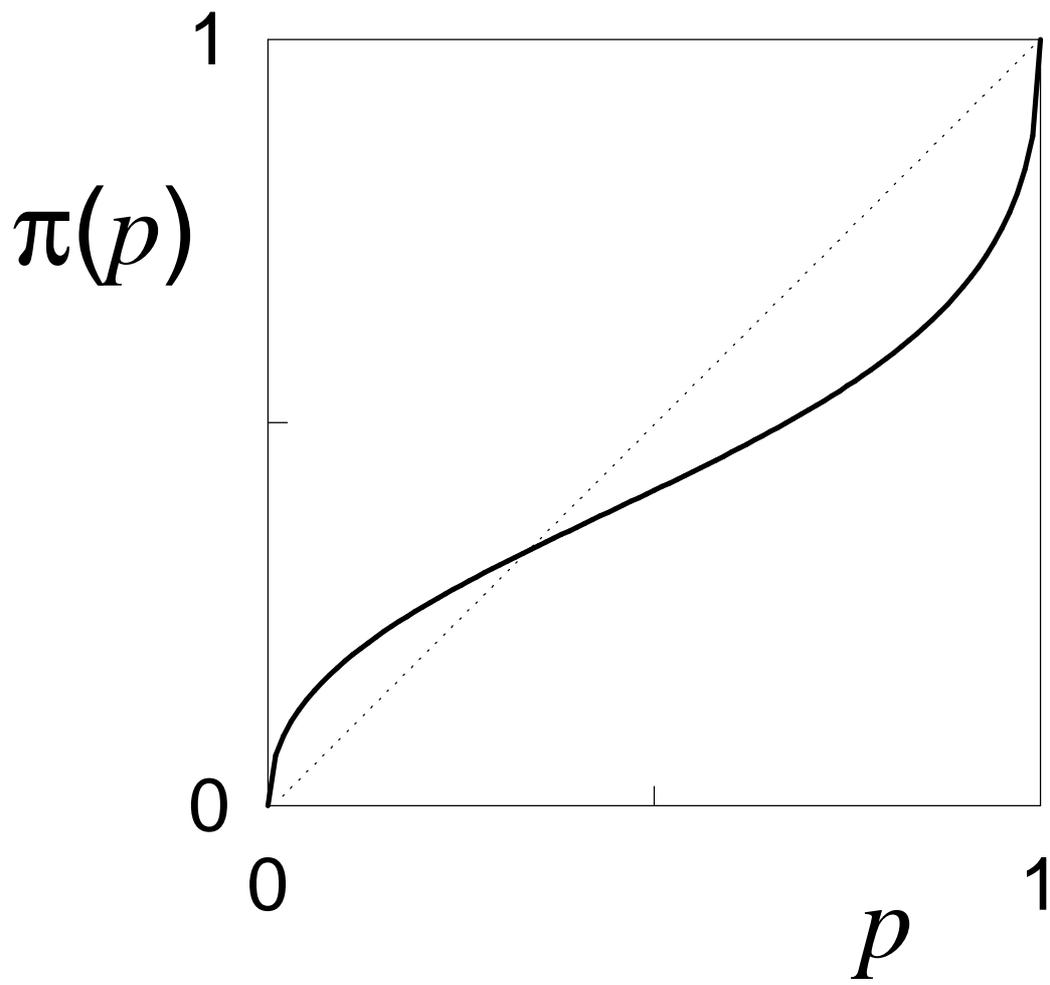
{\bf Figure 1}: Typical shape of the function $\pi(p)$.
The dotted line corresponds to the usual expectation value (EUT).
\\[2mm] \noindent
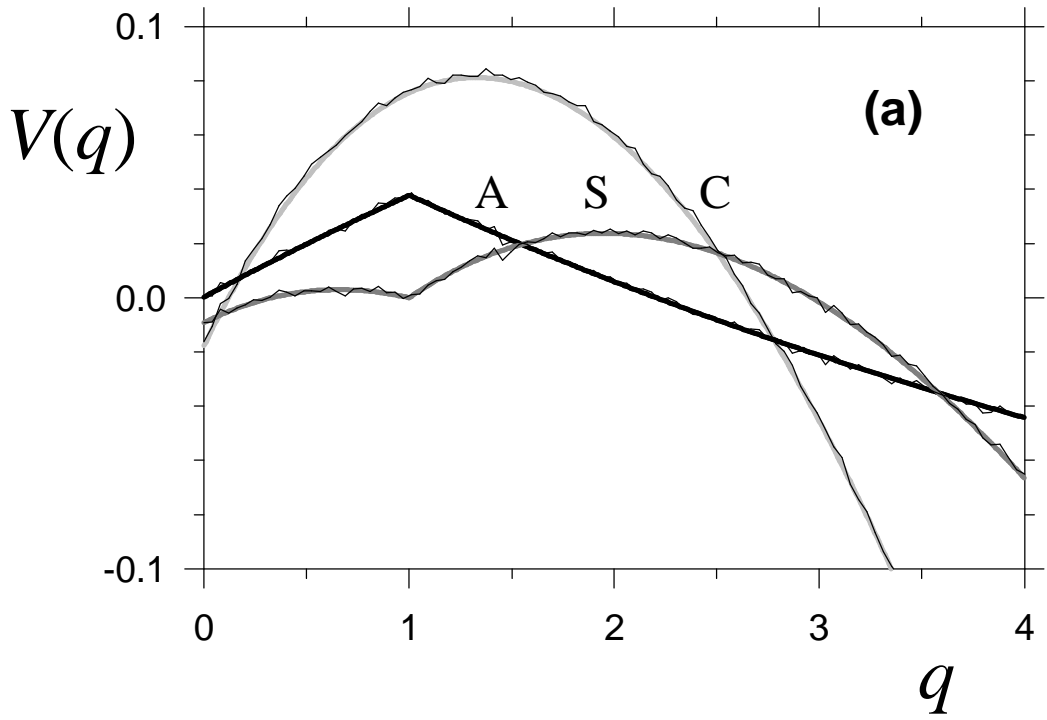
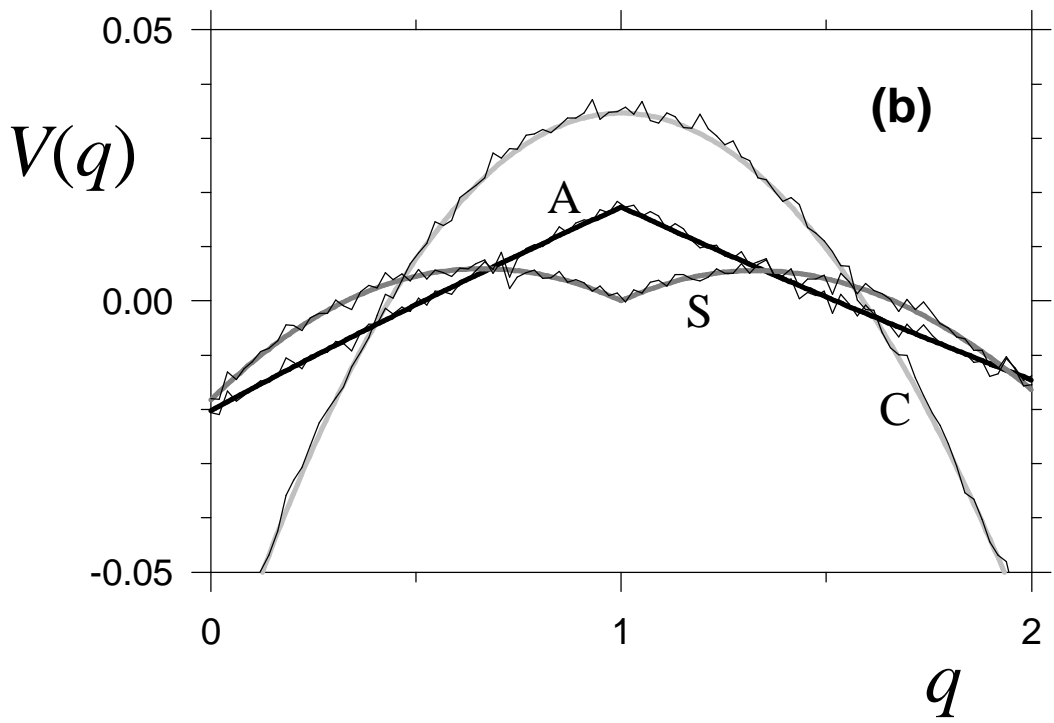
{\bf Figure 2}: $V(q)$ vs $q$, obtained from Eq. (\ref{Vq}), 
for cases A, S and C and two different initial
distributions of parameter $q$: uniform  in $[0,4]$ (a), and uniform in $[0,2]$
(b). 
Without debt restraints $V$ gives the average time evolution of assets:
$\overline{M}(q,t)-\overline{M}(q,0) =S t V(q)/N $.
Thin lines correspond to the average over $10^4$ numerical simulations
with $M(q,0)=M_o=1000$, $N=100$, $S=10$, $P=0.85$ and $t=10^4$.
\\[2mm] \noindent
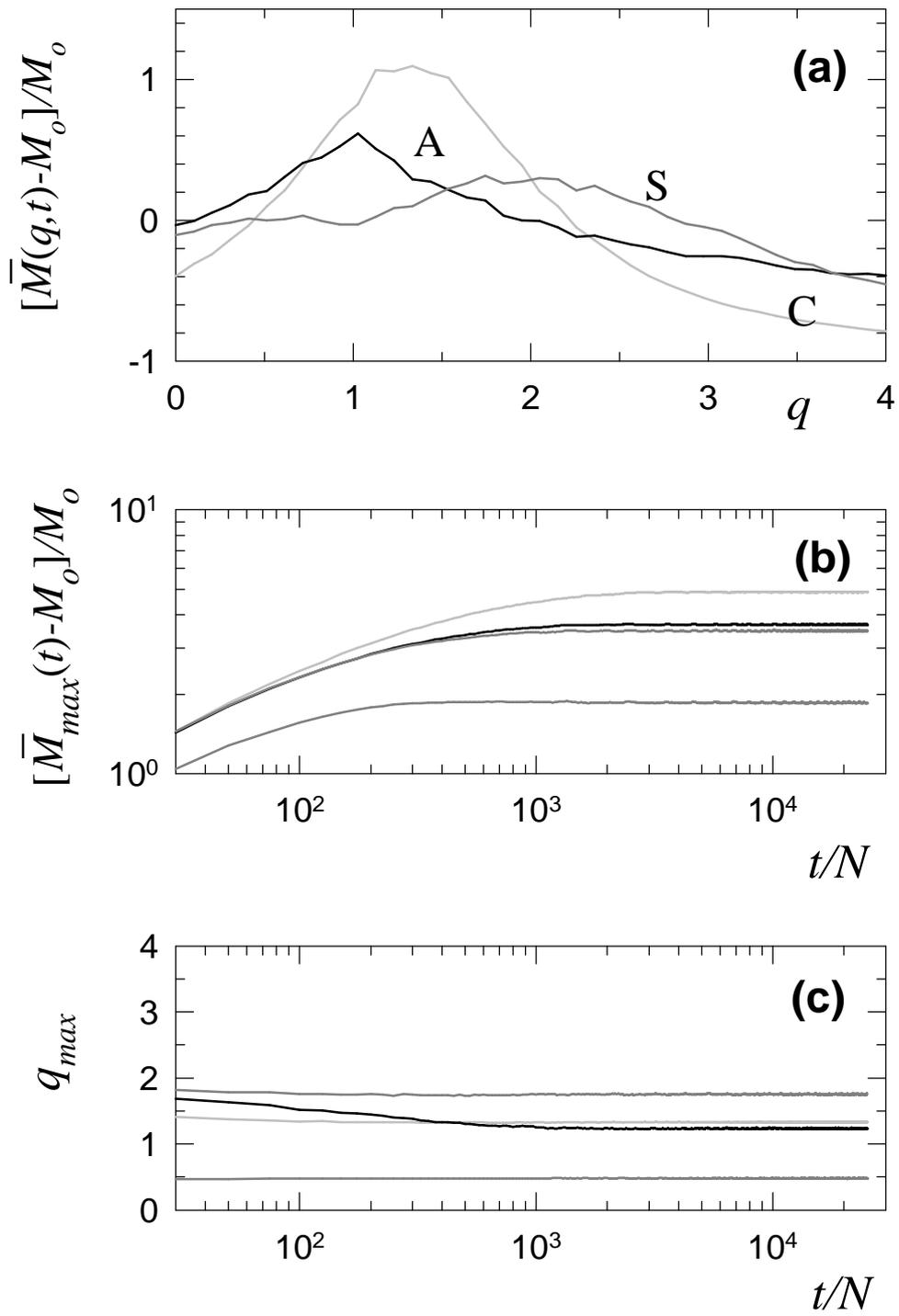
{\bf Figure 3}: Time  evolution of assets with indebtedness restraints of
kind (OR,*) (without exclusion) with threshold $M^*=100$.
(a)  ($\overline{M}(q,t)-M_o)/M_o$ vs $q$ at term
$t/N=25000$ when the steady state is already  attained, 
for cases A (black), S (dark gray) and C (light gray). 
Lines correspond to simulations averaged on $2\times 10^3$ experiments
with $M(q,0)=M_o=1000$, $N=40$, $S=100$ and $P=0.85$.  The plots do not depend on $N$
and depend on  $(M^*,M_o)$ only through their ratio.
Increasing $S$ shortens the time scale and increases fluctuation amplitude.
(b) $\overline{M}(q_{max},t)$ vs $t$ and 
(c) $q_{max}$ vs $t$ where $q_{max}$ maximizes $\overline{M}(q,t)$. 
For  case S, both maxima are plotted.
\\[2mm] \noindent
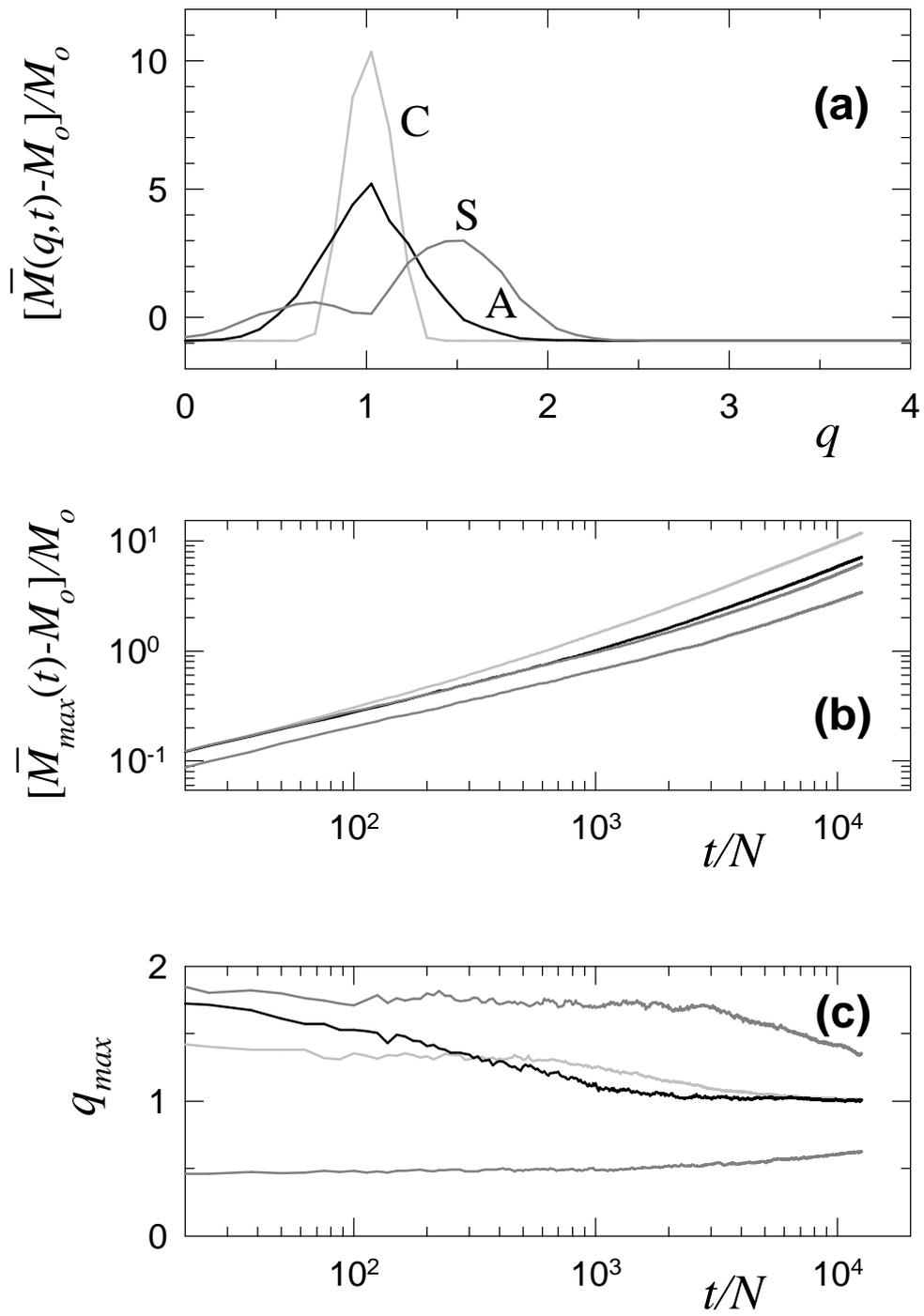
{\bf Figure 4}: Time evolution of assets with indebtedness restriction of
kind (PR,*) (with exclusion) with threshold $M^*=100$.
(a)  $( \overline{M}(q,t)-M_o)/M_o$ vs $q$ at term $t/N=12500$, 
for cases A (black), S (dark gray) and C (light gray). 
Lines correspond  to simulations averaged on $10^3$ experiments
with $M(q,0)=M_o=1000$, $N=40$, $S=10$ and $P=0.85$.
(b) $\overline{M}(q_{max},t)$ vs $t$ and 
(c) $q_{max}$ vs $t$ where $q_{max}$ maximizes $\overline{M}(q,t)$. 
For  case S, both maxima are plotted.

%\pagestyle{empty}

%\begin{figure}[h]
%\centerline{\mbox{\epsffile{figure1.ps}}}
%\end{figure}

%\begin{figure}[h]
%\centerline{\mbox{\epsffile{figure2.ps}}}
%\end{figure}

%\begin{figure}[h]
%\centerline{\mbox{\epsffile{figure3.ps}}}
%\end{figure}

%\begin{figure}[h]
%\centerline{\mbox{\epsffile{figure4.ps}}}
%\end{figure}

\end{document}